\documentclass[fleqn,usenatbib]{mnras}

\usepackage{newtxtext,newtxmath}

\usepackage[T1]{fontenc}

\DeclareRobustCommand{\VAN}[3]{#2}
\let\VANthebibliography\thebibliography
\def\thebibliography{\DeclareRobustCommand{\VAN}[3]{##3}\VANthebibliography}

\usepackage{graphicx}	
\usepackage{amsmath}	
\usepackage{tabularx}
\usepackage{float}
\usepackage{placeins}

\newcommand{\lbol}{L_{\rm bol}}

\newcommand{\new}{\color{black}} 
\newcommand{\mbh}{M_{\rm BH}}

\newlength{\standardparindent}
\newenvironment{minipdef}[2]{\makebox[#1]{#2\ \hfill}%
  \begin{minipage}[t]{\dimexpr\linewidth-#1}%
  \setlength{\parindent}{\standardparindent}\noindent\ignorespaces}%
{\end{minipage}}

\title[Quasar Time Dilation Revisited]{
Revisiting The Cosmological Time Dilation of Distant Quasars: \\
Influence of Source Properties and Evolution}

\author[Brewer, Lewis \& Li]{
Brendon J. Brewer,$^{1}$\thanks{E-mail: bj.brewer@auckland.ac.nz}
Geraint F. Lewis,$^{2}$
and Yuan (Cher) Li$^{1}$
\\
$^{1}$Department of Statistics, The University of Auckland, Private Bag 92019, Auckland 1142, New Zealand\\
$^{2}$Sydney Institute for Astronomy, School of Physics, A28, The University of Sydney, NSW 2006, Australia\\
}

\date{}

\pubyear{\the\year}

\begin{document}
\label{firstpage}
\pagerange{\pageref{firstpage}--\pageref{lastpage}}
\maketitle

\begin{abstract} 
After decades of searching, cosmological time dilation was recently identified in the timescale of variability seen in distant quasars. Here, we expand on the previous analysis to disentangle this cosmological signal from the influence of the properties of the source population, specifically the quasar bolometric luminosity and the rest-frame emission wavelength at which the variability was observed. Furthermore, we consider the potential influence of the evolution of the quasar population over cosmic time. We find that a significant intrinsic scatter of
$0.288 \pm 0.021$ dex in the variability timescales,
which was not considered in the previous analysis, is favoured by the data. This slightly increases the uncertainty in the results. However, the expected cosmological dependence of the variability timescales is confirmed to be robust to changes in the underlying
assumptions. We find that the variability timescales increase
smoothly with both
wavelength and bolometric luminosity, {\new and that
black hole mass has no effect on the variability timescale once rest wavelength
and bolometric luminosity are accounted for}. Moreover,
if the standard cosmological model is correct, governed by relativistic expansion, we also
find very little cosmological evolution in the intrinsic variability timescales of distant quasars.
\end{abstract}




\section{Introduction}
The dilation of time has been a central aspect of the theory of relativity since its inception \citep{1905AnP...322..891E,Minkowski1908}. In the early days of cosmology, where the source of cosmological redshift was still under discussion, \citet{1939ApJ....90..634W} suggested that the timescale of the brightening and fading of distant supernovae could be used to distinguish a relativistic origin from a tired light model where photons lose energy due to some other mechanism. But like many probes of cosmology, the inhomogeneity of supernovae explosions ensured that their use as a cosmological clock would be challenging. By the close of the twentieth century, significant effort had been expended on calibrating a subset of supernovae, Type Ia, to use them as cosmological probes \citep[see][]{2022NCimR..45..549C}. In calibrating the dispersion of supernovae properties, it was found that the light curve duration could be used as a standardised tick, allowing the detection of the expected cosmological time dilation signal \citep{2001ApJ...558..359G,2005ApJ...626L..11F,2008ApJ...682..724B,2024MNRAS.533.3365W}.

The detection of the time dilation in other cosmological sources, on the other hand, has proven more difficult. For example, quasars can show significant variability over a range of wavelengths and timescales. However, searches for the cosmological time dilation signal in samples of quasars observed over decades failed to yield the expected stretching of timescales at higher redshifts, leading to the suggestion that quasar variability may not be intrinsic and might be due to some intermediate influence, namely microlensing by a cosmological distribution of black holes \citep[e.g.][]{1993Natur.366..242H,1997ApJ...482L...5H,2001ApJ...553L..97H,2010MNRAS.405.1940H,2022MNRAS.512.5706H}.  

Recently, \citet{2022MNRAS.514..164S}
presented a sample of 190 quasars, originally identified in the Sloan Digital Sky Survey and monitored in multiple bands for two decades. Unlike previous samples, these quasars were monitored for extended periods in identical bands, although it should be noted that a range of different observing facilities were employed over the entire observing period, resulting in significant gaps in the data. By considering the underlying quasar variability as a damped random walk (DRW), which naturally has a timescale parameter, \citet{2022MNRAS.514..164S} were able to infer characteristic timescales to the light curves in each of the
three wavebands -- $g$, $r$, and $i$. These inferred timescales, with their
associated uncertainties, were made publicly available.
\citet[][LB23 hereafter]{lewis2023detection} employed this timescale as a tick of a quasar clock and, in grouping quasars by their bolometric luminosity and the rest-frame emission wavelength, searched for a cosmological signal of $\Delta t_{\rm obs} = \Delta t_{\rm int} ( 1 + z )^n$, where $\Delta t_{int}$ is the intrinsic valiability timescale, $\Delta t_{\rm obs}$ is the observed variability timescale and $n=1$ for Friedmann-Lemaitre-Robertson-Walker (FLRW) cosmologies. They found $n=1.28^{+0.28}_{-0.29}$, encompassing the expected cosmological signal, but with a possible offset that they suggested was due to variation across the source population and potentially due to evolution with redshift.
In this paper, we set out to revisit the results of LB23 with
different modelling assumptions to test whether the result is robust.
The assumptions in this current study are more akin to standard regression models,
allowing us to find (potentially) a simple relation between the
variability timescales, the properties of the source, and the redshift.
We also consider modifying the likelihood function to account for the
asymmetric error bars on the measured timescales.
Our primary goal is to test whether the inferred value of $n$,
and the overall conclusions of LB23 are significantly
affected by these changes in the modelling assumptions.

The layout of this paper is as follows: the modelling assumptions are given in Section~\ref{sec:assumptions}, with the results (of both parameter estimation and model selection) appearing in Section~\ref{sec:results}. We conclude in Section~\ref{sec:conclusions}.

\section{Modelling Assumptions}\label{sec:assumptions}
We follow the general approach presented in LB23 by assuming that quasar variability possesses an intrinsic timescale which is a function of the quasar bolometric luminosity and the rest-frame emission wavelength of the observed variability. {\new Throughout this study, we ignored the given error bars
on the bolometric luminosity values, as they are very small compared to
the dispersion of the bolometric luminosity measurements. Specifically,
the mean errorbar on $\log_{10}(\lbol/\textnormal{(erg/s)})$ is 0.018, but the standard deviation of
all the $\log_{10}(\lbol/\textnormal{(erg/s)})$ values is 0.450.}

In the standard cosmology picture, the observed variability timescales result from combining the intrinsic timescales with the cosmological time dilation term that depends on redshift. However, this dependence is potentially different for alternative cosmologies; for example, in a tired light cosmology, there will be no dependence on redshift, and the intrinsic and observed variability would be identical. 

\begin{table}
\begin{tabular}{l|l|l}
Parameter & Meaning & Prior \\
\hline
$\beta_0$ & Baseline level for log-timescale & Uniform(-10, 10) \\
$\beta_1$ & Rate  of increase with log-wavelength & Uniform(-10, 10) \\
$\beta_2$ & Rate of increase with log-luminosity & Uniform(-10, 10) \\
$\beta_{12}$ & Nonlinear term & Uniform(-1, 1) \\
$n$ & Redshift dependence & Uniform(-1, 4) \\
$\sigma$ & Intrinsic scatter & Uniform(0, 1) \\
$\{c_i\}_{i=1}^{190}$ & Quasar-specific offset & Normal$\left(0, \sigma^2\right)$
\end{tabular}
\caption{The prior distributions used in the analysis.
All parameters are dimensionless and refer to base-10 logarithms
of variability timescales.\label{tab:priors}}
\end{table}

Three variability timescales are available for each of the 190
quasars in the sample, corresponding to the three different
wavebands of the observations.
The rest-frame wavelengths of the observations are given by
\begin{equation}
\lambda_g = \frac{4720 \mathring{\mathrm{A}}}{1+z} \ \ \ \ \ \ \ \ \         
\lambda_r = \frac{6415 \mathring{\mathrm{A}}}{1+z} \ \ \ \ \ \ \ \ \ 
\lambda_i = \frac{7835 \mathring{\mathrm{A}}}{1+z}
\end{equation}
where $z$ is the redshift of the quasar.
Each timescale acts as a `data point' in our analysis, so there
are $190 \times 3 = 570$ data points available.

LB23's implementation split the data points into twelve bins based on their characteristic properties (See Figure~2 in LB23). With this, the variation within each bin was considered a combination of an intrinsic variability timescale (one free parameter for each bin) coupled with a cosmological influence for the form $(1+z)^n$. In the following, we aim to simplify this approach by considering a continuous function for the intrinsic variability over the sample, removing the need for binning, and considering an intrinsic scatter in the properties of the variability timescale. Removing the need for binning also
allows us to consider all the data points, whereas a few observations were excluded in LB23's analysis (those that fell outside all of the 12 bins).

We now describe the mathematical model employed in this current study.
Denoting the bolometric luminosity of a quasar by $\lbol$, the rest wavelength
of an observation (data point) by $\lambda$, and the variability timescale by $\tau$,
we first define their logarithms as primed quantities
to simplify the notation:
\begin{align}
\lbol' &= \log_{10}\left(\lbol/(\textnormal{erg/s})\right) \\
\lambda' &= \log_{10}(\lambda/{\mathring{\mathrm{A}}}) \\
\tau' &= \log_{10}(\tau/\textnormal{days}) \\
{\new z'} & {\new =\log_{10}(1 + z)}.
\end{align}
The equation for the proposed regression surface, which gives the {\new expected value of the }variability
timescale $\tau'$ as a function of the proposed explanatory variables
$\lbol'$ and $\lambda'$, is given by
\begin{align}
{\new T} &= \beta_0 + \beta_1 \left(\lambda' - \widebar{\lambda'}\right)
 + \beta_2 \left(\lbol' - \widebar{\lbol'}\right) \notag\\
 &\quad\quad\quad
 + \beta_{12} \left(\lambda' - \widebar{\lambda'}\right)\left(\lbol' - \widebar{\lbol'}\right) \notag\\
 &\quad\quad\quad
 + {\new n\left(z' - \widebar{z'}\right)}.
\label{eqn:surface}
\end{align}
This representation replaces the 12 free parameters (one for each bin) in the
LB23 analysis.

\begin{table}
\begin{tabular}{l|l}
Parameter & Result (post. mean $\pm$ post. sd.)\\
\hline
$\beta_0$ & 3.377 $\pm$ 0.026 \\
$\beta_1$ & 0.96 $\pm$ 0.12 \\
$\beta_2$ & 0.232 $\pm$ 0.084 \\
$\beta_{12}$ & 0.46 $\pm$ 0.23 \\
$n$ & 1.14 $\pm$ 0.34 \\
$\sigma$ & 0.289 $\pm$ 0.021 
\end{tabular}
\caption{Posterior summaries for the model parameters.
These are rounded to two significant figures in the uncertainty,
and the same precision in the point estimate.\label{tab:estimates}}
\end{table}

This parametric form has the following parameters:
\begin{itemize}
\item  \begin{minipdef}{0.65cm}{$\beta_0$:}
    Baseline level for the log-timescale.
  \end{minipdef}
\vspace{-2mm}

\item  \begin{minipdef}{0.65cm}{$\beta_1$:}
    Relation between timescale and wavelength.
  \end{minipdef}
\vspace{-2mm}

\item  \begin{minipdef}{0.65cm}{$\beta_2$:}
    Relation between timescale and luminosity.
  \end{minipdef}
\vspace{-2mm}

\item  \begin{minipdef}{0.65cm}{$\beta_{12}$:}
    Nonlinear cross-term of regression surface.
  \end{minipdef}
\vspace{-2mm}

\item  \begin{minipdef}{0.65cm}{$n$:}
    Cosmological dependence term.
  \end{minipdef}
\end{itemize}
The values $\widebar{\lambda'}$, $\widebar{\lbol'}$,
{\new and $\widebar{z'}$}
are the arithmetic means of
the observed values of these quantities. Subtracting these in the expression
for the regression surface has two advantages. Firstly, the joint
posterior distribution for these parameters tends to be less correlated
when this is done, enhancing the computational efficiency of the analysis.
Secondly, the interpretation of the parameters (especially
$\beta_0$) is more straightforward
\citep{cohen2013applied}, easing the choice of prior distributions.
When the means are subtracted,
$\beta_0$ becomes simply the log-timescale corresponding to a typical
value of bolometric luminosity, rest wavelength, {\new and redshift} rather than the value
when the explanatory variables are zero, which is physically impossible.

\begin{figure*}
\centering
\includegraphics[width=1.0\textwidth]{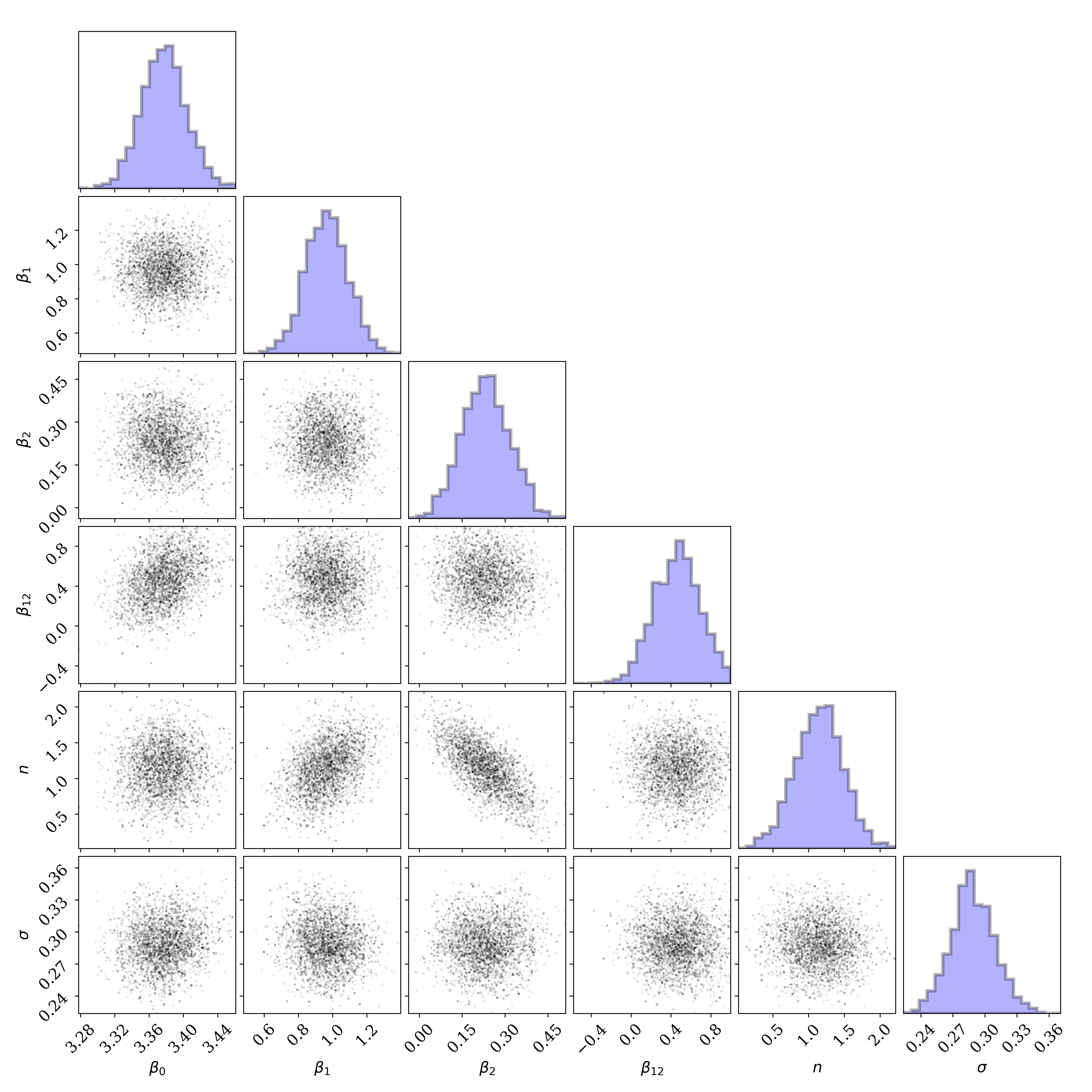}
\caption{A corner plot \citep{corner} of the posterior distribution for the
parameters. {\new The most substantial dependence is a
negative correlation between $\beta_2$ and $n$}.\label{fig:corner}}
\end{figure*}

\subsection{Likelihood Function}
In this analysis, we use a different form of the likelihood function from that considered by LB23. The original raw data consists of time series
observations \citep{2022MNRAS.514..164S}. However, it is prohibitive
to work with the time series directly. Instead, we use the inferred DRW timescales, which are
provided with 68\% non-symmetric credible intervals. Our analysis used this to build an asymmetric double exponential likelihood function.
We cannot easily treat these measurements
as `data' by defining a probability distribution for the data
given the parameters. Instead, we assume that the measurements provide an asymmetric likelihood function, which we can evaluate as a function of our model parameters. We form the likelihood function by assuming it is proportional to
a probability distribution that agrees with the given uncertainty quantiles.

In LB23, a skew-normal distribution was used for this purpose, but we chose an asymmetric double-exponential distribution in the present study.
This is for two reasons: (i) it greatly speeds up the computation, and (ii) we can choose the double-exponential parameters to match the log-timescale quantiles exactly rather than approximately as LB23 had to do with the
skew-normal distribution. The formula for the probability density function of
an asymmetric double-exponential is
\begin{align}
f(x) &= \left\{
    \begin{array}{lr}
    \frac{1}{2\ell_{\rm left}}\exp\left(\frac{x-\mu}{\ell_{\rm left}} \right), & x < \mu \\
    \frac{1}{2\ell_{\rm right}}\exp\left(-\left[\frac{x-\mu}{\ell_{\rm right}}\right]\right),
                                & x \geq \mu \\
    \end{array}
    \right.,\label{eqn:exponential}
\end{align}
which depends on parameters $\left(\mu, \ell_{\rm left}, \ell_{\rm right}\right)$. These are the median value, the length scale on the
left of the distribution, and the length scale on the right of the
distribution respectively.
For each log-timescale observation, we can
set the values of these three parameters from the given timescale quantiles,
using formulae given in Appendix~\ref{sec:exponential}.

{\new
Overall, the likelihood function, given the data
$\{\tau'_{1, 1}, .., \tau'_{1, 3}, ..., \tau'_{190, 3}\}$
is given by
\begin{align}
p(\boldsymbol{\tau} \,|\, \boldsymbol{\theta})
    &= \prod_{i=1}^{190}
    \prod_{j=1}^{3}
            f\left(T(\boldsymbol{\theta}) + c_i - \tau'_{i, j}\right)
\end{align}
where $\boldsymbol{\theta}$ denotes all parameters,
$T$ is given by Equation~\ref{eqn:surface}, and $f()$
is the function defined in Equation~\ref{eqn:exponential}.
}

\subsection{Intrinsic Scatter}
Up to now, the assumptions
(and those of LB23) imply that the true variability timescales $\tau'$( if
we could measure them with great precision) could be predicted exactly from
$\lambda'$ and $\lbol'$, an implausible situation.
To account for the fact that individual quasars may depart from the
exact relationship specified in Equation~\ref{eqn:surface}, we allow
each quasar to have its offset parameter $c_i$, describing its
departure from the relation given in Equation~\ref{eqn:surface}.

The prior for the offset parameters $\{c_i\}$ is centred around zero,
with the typical magnitude of the offsets given by a
hyperparameter $\sigma$. The prior for $\sigma$
and the offsets $\{c_i\}$ is specified hierarchically,
as follows:
\begin{align}
\sigma &\sim \textnormal{Uniform}(0, 1) \\
c_i    &\sim \textnormal{Normal}\left(0, \sigma^2\right).
\end{align}
We set a small upper limit of 1 for $\sigma$ because we
are dealing with log timescales, and a value of 1 would
correspond to an intrinsic scatter of plus or minus one
entire order of magnitude.

Each of the offset parameters $\{c_i\}$, as well as the hyperparameter $\sigma$,
is explicitly included in the parameter space and explored
in the posterior sampling process, as marginalising over the
$\{c_i\}$ is analytically intractable
(this tends to be possible only when the sampling distribution
for the data is Gaussian).
The hyperparameter $\sigma$ is known as the intrinsic scatter
or intrinsic dispersion, and describes the degree to which
individual quasars depart from the relation in Equation~\ref{eqn:surface}.

\subsection{Prior Distributions}
The prior distributions we assigned for all 
196 unknown parameters
are given in Table~\ref{tab:priors}.
For simplicity, most of these were chosen to be uniform but with a limited range to rule out
wildly implausible values and to facilitate
basic Bayesian model comparison, which we perform in Section~\ref{sec:model_comparison}.
Since the quantities on both sides of the regression model
are logarithms (to base 10), allowing a range from -10 to 10 for the
coefficients is a generously wide range. We find later
(Section~\ref{sec:results}) that
the posterior distributions fall well within the prior range.
The flat prior for $n$ is equivalent to that used in LB23.

\subsection{Computation}
The analysis was implemented in C++ using DNest4 \citep{dnest4},
which implements
Diffusive Nested Sampling \citep{dns},
a variant of the Nested Sampling algorithm \citep{skilling}
that uses Markov Chain Monte Carlo to explore the parameter space.
Diffusive Nested Sampling is based on the Metropolis algorithm
\citep{metropolis}, and tends to perform well in higher dimensions
as long as the proposal distributions are well chosen. This is in
contrast to some other popular sampling approaches \citep[e.g.][]{emcee, multinest}, which are less effective in high dimensions
\citep{huijser2022properties, dittmann2024notes}.

The overall dimensionality of the parameter space is 196, including
the regression parameters, the cosmological dependence
$n$, the quasar-specific offset parameters, and their hyperparameter
(the intrinsic scatter $\sigma$). Useful samples from the posterior distribution
and a marginal likelihood estimate were obtained within several minutes on a standard laptop computer.
However, the final results of this paper were produced in the long run, taking about two hours on a standard laptop computer.

\section{Results}\label{sec:results}

\subsection{Parameter Estimates}
At this stage of the analysis, all of the parameters in the model for the regression surface (Equation~\ref{eqn:surface}) were treated as free and summaries of their posterior distributions 
are given in
Table~\ref{tab:estimates}. Since all of the marginal posterior distributions
were close to Gaussian, we have chosen to summarise them using the
posterior mean $\pm$ the posterior standard deviation.
A corner plot is shown in Figure~\ref{fig:corner}.
The most significant dependence in the corner plot is
between {\new $\beta_2$} (the baseline level of log-timescales)
and $n$ (the cosmological dependence), and this is the source
of most of the uncertainty about $n$ that remains.

The inferred value of $n$
is $1.14 \pm 0.34$, which is slightly lower than LB23's estimate
of $1.28_{-0.29}^{+0.28}$, and with a slightly higher uncertainty.
The larger uncertainty is largely due to the inclusion of intrinsic
scatter in our model. The anticipated value of $n=1$ is comfortably
contained within the credible interval. However, in Section~\ref{sec:model_comparison}, we perform a more formal model
comparison to test the hypothesis that $n=1$.

\subsection{Regression Surface}
\begin{figure}
    \centering
    \includegraphics[width=0.5\textwidth]{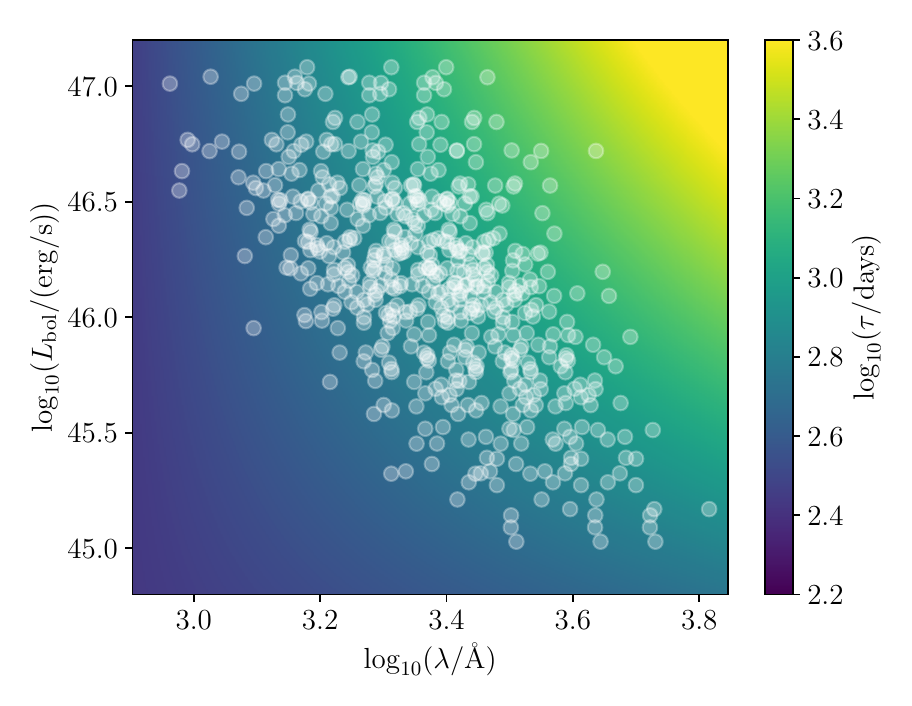}
    \caption{The posterior mean regression surface at $z=0$,
    showing the $\log_{10}$ of the intrinsic variation timescale as a function of
    wavelength and bolometric luminosity.
    Each point in the plot represents a measurement, so there
    are three points per quasar. The typical variation timescale
    {\new in the rest frame}
    is a little below $10^3$ days and increases smoothly as a function of rest wavelength
    and bolometric luminosity.\label{fig:surface}}
\end{figure}

The posterior mean of the regression surface
(Equation~\ref{eqn:surface}), excluding the cosmological
term, is shown in Figure~\ref{fig:surface}. This shows the expected value of
the ($\log_{10}$) variability timescale as a function of rest wavelength and
bolometric luminosity if the redshift were zero.
The relationship is approximately linear
(which can also be seen from the small inferred value of the nonlinearity
parameter $\beta_{12}=0.46 \pm 0.23$), and the timescale increases with both
$\lambda$ and $\lbol$.
The increase with $\log_{10}(\lambda/\mathring{\mathrm{A}})$ is more pronounced per unit ($\beta_1 = 0.96 \pm 0.12$) than the increase with $\log_{10}(\lbol/(erg/s))$ ($\beta_2 = 0.232 \pm 0.084$).
Over the range of explanatory variables present in the dataset,
$\log_{10}(\lambda)$ contributes slightly more to the dispersion in timescales.
These results are broadly consistent with those obtained by \citet{kelly2009variations}
using a different sample of quasars and different modelling assumptions.

\subsection{Marginal Likelihoods and Model Comparison}\label{sec:model_comparison}
The marginal likelihood, also called the Bayesian evidence, is the prior
probability (or probability density)
of the data $D$ irrespective of the value of
any parameters $\theta$. It is given by
\begin{align}
    p(D) &= \int p(\theta)p(D \,|\, \theta) \, d\theta,
\end{align}
where $p(\theta)$ is the prior distribution, $p(D\,|\,\theta)$ is
the likelihood function and the integral is over the entire parameter space
\citep{o2004kendall, skilling}.
These values play the role of likelihood when computing the
posterior probability of a model compared to its alternatives.
Models are also often compared using Bayes Factors, ratios
of marginal likelihoods for two models. In this section, we consider
a range of variations on the model and test them against the version
of the model presented so far.
Unfortunately, we cannot compare our marginal likelihoods to LB23's marginal likelihood due to differences in the data caused by binning, which excluded several data points. Nevertheless, we present alternative versions of the model and their corresponding marginal likelihoods. 

\begin{table*}
\begin{tabular}{l|l|l|l}
Model name & Description & $\ln(Z)$ & Bayes Factor vs. $M_0$\\
\hline
$M_0$ & Main Model & $-203.50$ & 1 \\
$M_1$ & Additional quadratic terms & $-205.98$ & $0.083$ \\
$M_2$ & Extra intrinsic scatter parameters (three per quasar) & $-294.35$ & $3.5 \times 10^{-40}$ \\
$M_3$ & No intrinsic scatter at all & $-296.26$ & $5.2 \times 10^{-41}$ \\
$M_4$ & $\lambda/(1000\mathring{\mathrm{A}})$ replaces $\log_{10}(\lambda/\mathring{\mathrm{A}})$ in model equation & $-211.32$ & $4.0 \times 10^{-4}$ \\
$M_5$ & Additional evolution term given by $\beta_3 z$ & $-207.35$ & $0.021$ \\
$M_6$ & Additional evolution term given by $\beta_3 z$, but $n$ fixed to 1 & $-207.97$ & $0.011$ \\
$M_7$ & Relativistic Cosmologies \& No Evolution ($n=1$, $\beta_3=0$) & $-202.25$ & $3.5$ \\
{\new $M_8$} & {\new Black hole mass included} &  {\new $-207.40$} & {\new 0.020} \\
{\new $M_9$} & {\new Black hole mass included, bolometric luminosity removed} & {\new $-206.57$} & {\new $0.046$}
\end{tabular}
\caption{Marginal likelihoods for the various models considered
in this paper. The description column states how the model differs from the main model. Bayes Factors are presented relative to the main model.
The model with the most support from the data is $M_7$, which assumes no evolution in the behaviour of quasars and that standard cosmology applies.\label{tab:logzs}}
\end{table*}

\begin{figure}
\centering
\includegraphics[width=0.5\textwidth]{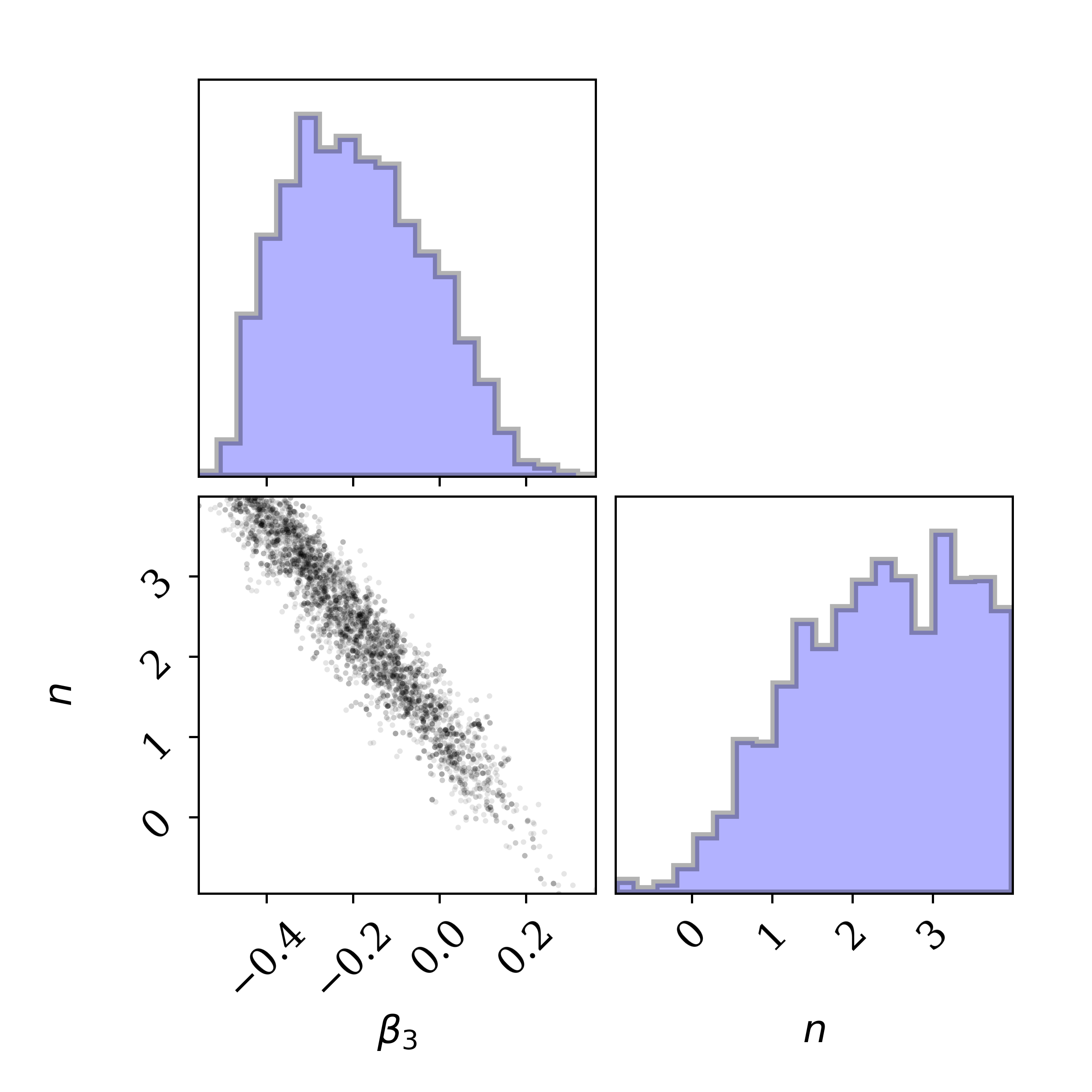}
\caption{The joint posterior distribution of $\beta_3$ (the coefficient of an evolution term), and $n$, the cosmological dependence. There is a very strong dependence between these two parameters. Taking a vertical slice at $\beta_3=0$ reproduces the main model's results. \label{fig:corner2}}
\end{figure}

\subsubsection{$M_0$: The Main Model}
Throughout this section, the model discussed thus far will be referred to as the {\em main model}, $M_0$. Table~\ref{tab:logzs} summarises the marginal likelihoods for all models considered alongside Bayes Factors relative to the main model.
The marginal likelihood of the main model was computed using DNest4, yielding an estimate of $\ln(Z) = -203.50$; note that DNest4 does not estimate uncertainties on $\ln(Z)$. 

\subsubsection{$M_1$: Additional Quadratic Terms}
Here we considered a modification of the regression surface through the
introduction of additional quadratic terms proportional to $\left(\lambda' - \widebar{\lambda'}\right)^2$ and $\left(\lbol' - \widebar{\lbol'}\right)^2$, incorporating two extra parameters for the coefficients of these terms.
{\new These parameters had Uniform$(-1, 1)$ priors, the same
as $\beta_{12}$.}
 This extension provides greater flexibility in the potential shape of the regression surface. The estimated marginal likelihood for this model was $-205.98$, lower than that of the main model. The inferred values for the two additional coefficients were close to zero, suggesting that these terms are unnecessary for describing the regression surface given the available data.

\subsubsection{$M_2$: Additional Intrinsic Scatter Parameters}
Here we examined the effect of altering the assumptions concerning intrinsic scatter. In the main model, one offset parameter is assigned per quasar, resulting in 190 such parameters. As an alternative, model $M_2$ introduces an offset parameter for each observation (i.e., a distinct offset for each of the three bands per quasar), resulting in $3 \times 190 = 570$ offset parameters. This model yielded a marginal likelihood of $\ln(Z) = -294.35$, substantially lower than that of the main model. The current data demonstrates that, for an individual quasar, the same variability is reflected across all the wavebands
(i.e., if the timescale is unusually long in one band, it will be in the other bands as well). 

\begin{figure}
\centering
\includegraphics[width=0.5\textwidth]{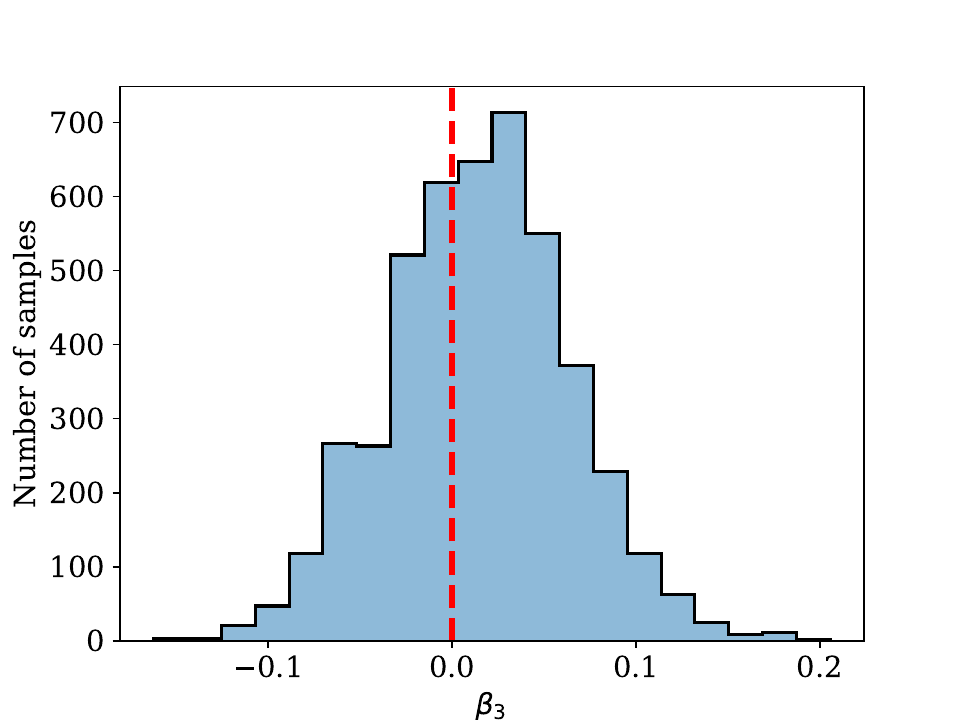}
\caption{The posterior distribution of $\beta_3$ (the coefficient of an evolution term), under model $M_6$. The parameter $n$ was fixed to 1, i.e., standard cosmology.
The inferred value of $\beta_3$ is small and consistent with zero
(vertical dashed line).\label{fig:beta3}}
\end{figure}

\subsubsection{$M_3$: Removal of Intrinsic Scatter}
We also explored the scenario in which the intrinsic scatter is entirely removed (i.e., $\sigma$ and all $\{c_i\}$ set to zero), referred to as model $M_3$. This model produced a marginal likelihood of $-296.26$, significantly lower than the main model, with the inferred value of  $n=1.27 \pm 0.25$. Among all the models considered in this work, this one most closely aligns with the assumptions of LB23. The inferred value for $n$, along with its uncertainty, confirms that the slightly larger uncertainty about $n$ under the main model is due to the inclusion of intrinsic scatter.

\subsubsection{$M_4$: Modification of the Regression Surface}
We also investigated a model $M_4$ in which the dependence of $\log_{10}(\tau)$ on rest wavelength uses $\lambda/(1000 \mathring{\mathrm{A}})$ in place of $\log_{10}(\lambda/\mathring{\mathrm{A}})$. The factor of 1000 here scales the explanatory variable similarly to before, thus removing the need for careful consideration of the prior widths for the coefficients $\beta_1$ and $\beta_{12}$ to ensure fair comparison between models. This modification allows for a slightly different set of regression surfaces. The marginal likelihood for this model was $\ln(Z)=-211.32$, slightly lower than that of the main model. The conclusions regarding $n$ remain similar to those drawn from the main model.

\subsubsection{$M_5$: Additional Evolution Term}
We also introduced an additional evolution term, $\beta_3 z$, intended to capture possible changes in quasar behaviour over cosmological time. This term resembles the time dilation term but features a slightly different functional form as a function of $z$. Since both terms describe trends with redshift, it is expected that disentangling the two possible causes of redshift dependence in the timescales will be challenging. A Uniform$(-10, 10)$ prior was applied for $\beta_3$. As anticipated, the posterior distribution showed a strong dependence between $\beta_3$ and $n$ (see Figure~\ref{fig:corner2}), where an increase in one is offset by a decrease in the other to fit the data. Thus, to draw conclusions about the cosmological timescale, strong assumptions about the lack of evolution are necessary. Conversely, to make strong conclusions about evolution, strong assumptions regarding cosmological time dilation, such as fixing $n=1$, must be made. The marginal likelihood for this model was slightly lower than that of the main model, with $\ln(Z)=-207.35$.

\subsubsection{$M_6$: Additional Evolution Term, with $n=1$}
Here, we retained the evolutionary term but with $n$ is fixed at 1 (i.e., assuming standard cosmological time dilation). The inference for $\beta_3$ in this case allows us to assess whether and how quasar variability evolves with cosmic time. The result was $\beta_3 = 0.014 \pm 0.049$, indicating very little evolution. The posterior distribution is shown in Figure~\ref{fig:beta3}. The marginal likelihood of this model was $\ln(Z) = -207.97$.

\subsubsection{$M_7$: Relativistic Cosmologies \& No Evolution}
In this model, we consider the case where $n=1$ and $\beta_3=0$. This represents an assumption that standard cosmology applies and that quasar behaviour shows no evolution. The marginal likelihood for this model was $-202.25$, the highest of all models considered, providing further evidence that the standard cosmological picture is correct and that there is no detectable evolution in quasar behaviour in the present dataset. {\new The parameter estimates from this model are given in
Table~\ref{tab:estimates2}. In most cases, these are similar
to the previous estimates.
}

\begin{table}
{\new
\begin{tabular}{l|l}
Parameter & Result (post. mean $\pm$ post. sd.)\\
\hline
$\beta_0$ & 3.371 $\pm$ 0.027 \\
$\beta_1$ & 0.94 $\pm$ 0.12 \\
$\beta_2$ & 0.252 $\pm$ 0.060 \\
$\beta_{12}$ & 0.44 $\pm$ 0.23 \\
$n$ & 1 \\
$\sigma$ & 0.290 $\pm$ 0.021
\end{tabular}
\caption{Posterior summaries for the model parameters
under model $M_7$, where $n$ is fixed to 1.
These are rounded to two significant figures in the uncertainty,
and the same precision in the point estimate.\label{tab:estimates2}}
}
\end{table}

{\new
\subsubsection{$M_8$: Black Hole Mass as an Explanatory Variable}
The \citet{2022MNRAS.514..164S} dataset includes estimated black hole masses
$\mbh$, with errorbars for the 190 quasars. Previous work has discovered
a correlation between $\mbh$ and $\tau$ \citep{burke2021characteristic}
over a very wide dynamic range of black hole masses from $10^4$ to
$10^{10} M_{\odot}$, (much wider than the range considered here).
To test the effect of black hole mass, we added $\log_{10}\left(\mbh/M_{\odot}\right)$ as an explanatory variable, with coefficient $\beta_4$.
For simplicity, we kept everything else the same as in the main model.

To account for the error bars on $\mbh$, which are significant, it is
necessary to include the true $\mbh$ values as nuisance parameters in
the analysis \citep{kelly2007some}, and to perform the regression against the true values
rather than the measured values. This also improves the clarity of the interpretation
of any resulting inferences. However, to perform model comparison,
we cannot treat the measured $\mbh$ values as additional data
(which would add extra terms to the likelihood function making it incomparable with previous analyses as the data has changed). Therefore, we must treat the $\mbh$ measurements as prior information
rather than as data. To do this, we first constructed the
prior distribution for the true black hole masses given the
measurements.

Letting the true log of the black hole mass of quasar $i$ be
\begin{align}
M_i' &= \log_{10}(\mbh{}_{,i}/M_\odot),
\end{align}
we assigned a prior distribution conditional on hyperparameters
$\mu_M$ and $\sigma_M$:
\begin{align}
M_i' \sim \textnormal{Normal}\left(\mu_M, \sigma_M^2\right).
\end{align}
We used the following likelihood
function for the observed/measured black hole masses:
\begin{align}
M_{i, \rm obs}' \sim \textnormal{Normal}\left(M_i', s_i^2\right)
\end{align}
where $s_i$ is the reported error bar on the log black
hole mass of quasar $i$.
We inferred the values of $\mu_M$ and $\sigma_M$
from the measured black hole masses, and then computed
point estimates of 8.92 and 0.40 respectively.
For simplicity, these were fixed for the rest of the
analysis, and formed part of the prior for the true
black hole masses (given the black hole mass measurements
but not the $\boldsymbol{\tau}$):
\begin{align}
p(M_i' \, | \, \boldsymbol{M}_{i, \rm obs})
    &\propto \exp\left(-\frac{1}{2\sigma_M^2}(M_i' - \mu_M)^2\right)
        \times \\
     &\quad\quad   \exp\left(-\frac{1}{2 s_i^2}(M_i' - M_{i, \rm obs})^2\right).
\end{align}

An extra term given by $\beta_4 M'$ was added to the regression
equation~\ref{eqn:surface}, and the coefficient $\beta_4$ was estimated along with all other parameters and the unknown true black hole masses. The prior assigned to $\beta_4$ was a Uniform$(-10, 10)$
distribution.
The estimated marginal likelihood of this model is $-207.40$, lower
than the main model, suggesting that black hole mass does not
provide any additional predictive power about the variability
timescale beyond what is already provided by $\lbol$ and
$\lambda$. This conclusion is supported by the inferred value
of $\beta_4 = 0.121 \pm 0.072$, which has significant overlap
with zero. This result is not in contradiction with
\citet{burke2021characteristic}, who looked at the correlation
between $\tau$ and black hole mass without including
$\lbol$.

We also experimented with adding nonlinear terms
involving the black hole mass, but none of the resultant
models was preferred over the main model or over $M_8$. These results
are not included in the paper.
}

{\new
\subsubsection{$M_9$: Black Hole Mass Included, Bolometric Luminosity Removed}
Since black hole mass is correlated with bolometric luminosity,
it is possible that including $\mbh$ as an explanatory
variable would remove the need for $\lbol$.
To test this, we implemented the model with black hole mass
included (as in the previous subsection) but without bolometric
luminosity. The result was a marginal likelihood of
$-206.57$, lower than the main model with a Bayes Factor of
0.046. This suggests a model including $\lbol$ but not
$\mbh$, i.e., the main model, is favoured.
}

\section{Conclusions}\label{sec:conclusions}
In this paper, we revisited the question of whether the cosmological time dilation signal can be detected in a sample of 190 quasars with time variability data. We refined the assumptions used by \citet{lewis2023detection} to align them more closely with standard regression modelling approaches, incorporating an intrinsic scatter term and modifying the form of the likelihood function. Our objective was to assess whether these changes would alter the overall conclusions and to explore the potential impact of evolution in the source properties over cosmic time.

Despite these adjustments, we still detect the cosmological signal (albeit with slightly increased uncertainty), finding that a cosmological dependence of the form $(1+z)^n$ yields $n=1.14 \pm 0.34$, consistent with the expectations from relativistic cosmologies. We compared our modelling assumptions against several alternatives and demonstrated that our main model was favoured over most others using a Bayesian model comparison approach. The only exception was a model that assumes $n=1$, which outperformed the main model. Additionally, assuming standard cosmology $(n=1)$, we also investigated the possibility of evolution in quasar variability timescales over cosmic time. We found that if such evolution exists, its magnitude must be small and consistent with zero. With the confirmation of the presence of the cosmological time dilation, we will have to await future large surveys of quasars to determine the presence of any timescale evolution or whether the physics of quasar variability is a constant across the life of the universe.

\section*{Acknowledgements}
GFL thanks the Royal Astronomical Society of New Zealand for partially funding his travel to Auckland where the writing of this paper began.
{\new We thank the anonymous referee for providing useful suggestions
to improve the analysis and the paper.}
\section*{Data Availability}
The data employed in this study was made publicly available by \citet{2022MNRAS.514..164S}. The code developed for this study,
along with the relevant subset of the data in a plain text format,
is available at
\url{https://github.com/eggplantbren/QuasarTimeDilation2}. The
code is available under the MIT Licence.


\bibliographystyle{mnras}
\bibliography{paper} 




\appendix

\section{Two-Sided Exponential Likelihood Details}\label{sec:exponential}

We need to be able to take the quantiles (16\%, 50\%, and 84\%) {\new for $\log_{10}(\tau)$} provided
by \citet{2022MNRAS.514..164S} and produce a likelihood function from them. {\new Traditionally, with symmetric errorbars, this would be
done using a Gaussian or normal distribution.
However, here we have significantly asymmetric errorbars.
To account for this, in the LB23 analysis, the mapping from the quantiles to a likelihood
function was done using the skew-normal distribution
\citep{o1976bayes}. However, it is not always possible to fit
a distribution to the quantiles exactly with this family of
distributions. Also, the narrow tails of this distribution
may lead to overconfident inferences. Therefore we chose
to construct a likelihood function that is proportional to
a two-sided exponential distribution.}

Technically, the given quantiles refer to posterior distributions rather than likelihood functions,
but the two are proportional provided the prior over $\log \tau$ was
approximately uniform in the original analysis of the time series data,
{\new which is true in this case}.
The idea is that it would be useful for analyses to provide
likelihood functions as output, rather than (or in addition to) posterior
distributions, is explored by \citet{2018arXiv180407766H}.

\begin{figure}
    \centering
    \includegraphics[width=0.5\textwidth]{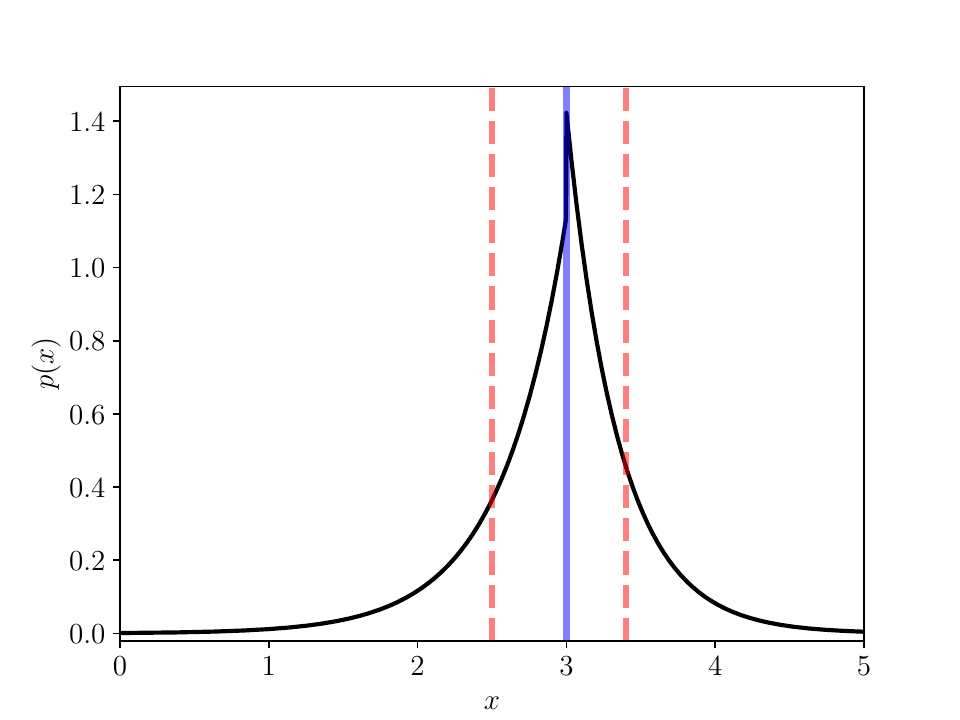}
    \caption{An example of an asymmetric exponential distribution
        (proportional to our likelihood function) with quantiles
        set at $x=(2.5, 3.0, 3.4)$.
    \label{fig:exponential}}
\end{figure}

Consider a probability density function composed of a mixture of two parts: a regular exponential
distribution on the right-hand side, and a reversed exponential distribution on
the left-hand side. Suppose it is centred at a median value $x=\mu$, with a scale length of $\ell_{\rm left}$ on the left and $\ell_{\rm right}$ on the right. Immediately, the median value $\mu$ can be set from the
given 50\% quantile.
The overall probability density is given by
\begin{align}
p(x) &= \left\{
    \begin{array}{lr}
    \frac{1}{2\ell_{\rm left}}\exp\left(\frac{x-\mu}{\ell_{\rm left}} \right), & x < \mu \\
    \frac{1}{2\ell_{\rm right}}\exp\left(-\left[\frac{x-\mu}{\ell_{\rm right}}\right]\right),
                                & x \geq \mu \\
    \end{array}
    \right.
\end{align}
An example of this probability density is given in Figure~\ref{fig:exponential}.
With $\mu$ set to the 50\% quantile, we
now need to find the formulas for setting $\ell_{\rm left}$ and $\ell_{\rm right}$ based on
the 16\% and 84\% quantiles.

For a standard Exponential distribution
with a scale parameter of 1, the inverse cumulative distribution function is
\begin{align}
F^{-1}(u) &= -\log(1 - u).
\end{align}
If we find the 68\% quantile of this distribution, it will
correspond to the position of the 68\% quantile of the
double exponential distribution as well.
This occurs at
\begin{align}
x &= -\log(1 - 0.68) \\
  &\approx 1.1394.
\end{align}
Therefore, 68\% of the mass of the double exponential distribution is
contained between $\mu-1.1394\ell_{\rm left}$ and $\mu + 1.1394\ell_{\rm right}$.
If we are given the credible interval $[x_l, x_r]$ we can solve for the
$\ell$ values using
\begin{align}
\ell_{\rm left} &= \frac{\mu - x_l}{1.1394} \\
\ell_{\rm right} &= \frac{x_r - \mu}{1.1394}.
\end{align}
This completes the process of computing the
likelihood parameters $\left(\mu, \ell_{\rm left}, \ell_{\rm right}\right)$
from the given quantiles.

We note that the Laplace (biexponential) distribution is a special
case of this distribution when $\ell_{\rm left} = \ell_{\rm right}$,
i.e., when the error bars are symmetric. Furthermore,
while it is not the maximum entropy distribution
\citep{jaynes2003probability} given only
quantile constraints (which is impractical), the Laplace
distribution does have higher entropy than the Gaussian
with equivalent quantiles.

{\new As a further check on our choice of likelihood function,
we implemented an alternative analysis with a two-sided gaussian
function instead. The marginal likelihood with this alternative
likelihood function was $-204.38$, slightly lower than for the
exponential. All other conclusions drawn from the results remained
similar to the main analysis conducted in this paper, except
the inferred value of $n$ reduced slightly to $1.10 \pm 0.32$.

A symmetric gaussian likelihood, with standard deviation given
by the geometric mean of the upper and lower error bars,
yields a low marginal likelihood of $-220.02$. The resulting
parameter estimates from this model are virtually identical
to the results from the main model.
}


\bsp	
\label{lastpage}
\end{document}